\documentclass[aps,nofootinbib,superscriptaddress,preprint,showpacs]{revtex4}
\usepackage{graphicx}
\usepackage{amssymb}
\usepackage{xcolor}
\begin{document}

\title{Compression modulus and symmetry energy of nuclear matter with KIDS density functional}
\author{Hana Gil}
\affiliation{Center for Extreme Nuclear Matter, Korea University, Seoul 02841, Korea}

\author{Chang Ho Hyun}
\email{hch@daegu.ac.kr}
\affiliation{Center for Extreme Nuclear Matter, Korea University, Seoul 02841, Korea}
\affiliation{Department of Physics Education, Daegu University, Gyeongsan 38453, Korea}

\date{\today}

\begin{abstract}
Equation of state of dense nuclear matter is explored in the KIDS density functional theory.
Parameters of the equation of state which are coefficients of the energy density expanded in powers of
$(\rho - \rho_0)/3\rho_0$ where $\rho$ is the nuclear matter density and $\rho_0$ is its density at saturation are 
constrained by using both nuclear data and the mass-radius relation of the neutron star determined from the modern astronomy.
We find that the combination of both data can reduce the uncertainty in the equation of state parameters significantly.
We confirm that the newly constrained parameters reproduce the basic properties of spherical magic nuclei with high accuracy.
Neutron drip lines, on the other hand, show non-negligible dependence on the uncertainty of the nuclear symmetry energy.
\end{abstract}
\maketitle

\section{Introduction}

The era of precision observation of the neutron star, and the multi-messenger astronomy and astrophysics has begun.
One of the most fundamental questions in the neutron star physics is what is the correct equation of state (EoS) of
nuclear matter at densities below and above the saturation density ($\rho_0$). 
Equation of state is determined by the energy of a particle and pressure which is gradient of energy with respect to density.
Basic properties of the neutron star such as mass, size, density profile, tidal deformability, species of particles that
consist of the neutron star matter, and their fractions are directly correlated to the EoS.
Neutron star cooling is critically dependent on the fraction of particles, so the EoS also plays an essential role in the thermal
evolution of neutron stars.

Energy of a particle in nuclear matter is conventionally expanded around the saturation density in terms of $x = (\rho - \rho_0)/3 \rho_0$ as
\begin{eqnarray}
E(\rho,\, \delta) &=& E(\rho) + S(\rho) \delta^2 + O(\delta^4), \\
E(\rho) &=& E_B + \frac{1}{2} K_0 x^2 + O(x^3), \label{eq:sat} \\
S(\rho) &=& J + L x + \frac{1}{2} K_{\rm sym} x^2 + \frac{1}{6} Q_{\rm sym} x^3 + O(x^4). \label{eq:sym}
\end{eqnarray}
Neutron-proton asymmetry is accounted by $\delta = (\rho_n - \rho_p)/\rho$ where $\rho_n$ and $\rho_p$ are the 
density of the neutron and the proton, respectively.
Most of the nuclear structure models fit their model parameters to the nuclear data, and 
the constants and coefficients in Eqs.(\ref{eq:sat}, \ref{eq:sym}) which determine the EoS of nuclear matter are 
obtained as results of the fitting. 
Among the constants and coefficients in Eqs.(\ref{eq:sat}, \ref{eq:sym}),
saturation density $\rho_0$ and the binding energy $E_B$ are determined highly model-independently,
and the values are converged to $\rho_0=0.16$ fm$^{-3}$ and $E_B = 16$ MeV.
On the other hand, compression modulus $K_0$ that characterizes the density dependence of EoS in the symmetric nuclear matter
is uncertain more than $\rho_0$ and $E_B$.
For example, collection of 240 Skyrme force models shows that the models predict the $K_0$ value in the range $200-400$ MeV \cite{dutra}.
Experimentally constrained range is reduced to $200-260$ MeV, but it is still uncertain much more than $\rho_0$ and $E_B$.
Since the average density of a canonical neutron star, a star of mass $1.4 M_\odot$ is more than $2\rho_0$,
role of $K_0$ in the neutron star matter might be non-negligible and therefore it is needed to reduce the uncertainty of $K_0$
in order to have more exact EoS of the neutron star matter.
Situation is no better than $K_0$ for the symmetry energy $S(\rho)$.
Ranges of $J$, and $L$ constrained from experiments are $30-35$ MeV and $40-76$ MeV, respectively \cite{dutra}.
For $K_{\rm sym}$ experimental access is still very limited, and the value is guessed from relations with $L$ or $3J-L$.
Properties of neutron rich systems ($\delta \simeq 1$) are sensitive to the behavior of the symmetry energy.
$J$, $L$ and $K_{\rm sym}$ being siginificantly uncertain, large uncertainty in the EoS of neutron stars is unavoidable.

Goal of this work is to reduce the uncertainty of $K_0$, $J$ and $L$ so that we have more reliable EoS for the neutron star
matter at densities up to $3\rho_0$ or $4\rho_0$ below which nucleons may be the dominant constituents of the matter.
It has been shown that if the energy density of nuclear matter is expanded in powers of the Fermi momentum of the nucleon,
7 terms are sufficient for the correct description of EoS at densities relevant to nuclei and the center of neutron stars 
\cite{kidsnm, kidsnuclei1, kidsnuclei2}.
If higher order terms are added to the basic 7 terms, additional terms give vanishing contribution up to $\rho \simeq 3\rho_0$,
and their effects are marginal at densities larger than $3\rho_0$.
Following this observation, we describe the EoS of infinite nuclear matter with 7 density-dependent terms.
Seven parameters in the functional are fixed to the values of $\rho_0$, $E_B$, $K_0$, $J$, $L$, $K_{\rm sym}$, and $Q_{\rm sym}$.
After the 7 nuclear matter parameters determined, the model is applied to nuclei.
Energy per particle ($E/A$) and the charge radii ($R_c$) of spherical magic nuclei $^{40}$Ca, $^{48}$Ca and $^{208}$Pb 
are reproduced by adjusting two additional parameters that account for the gradient of density and spin-orbit interactions \cite{polb}.
At this stage the number of model parameters is increased to 9, but we don't need to introduce additional parameters any more.
By varying $K_0$, $J$, $L$ and $K_{\rm sym}$ values, $\chi^2$ values are calculated for $E/A$ and $R_c$ of $^{40}$Ca, $^{48}$Ca and $^{208}$Pb.
We find that nuclear data are most accurately reproduced in the range $K_0 = 230-250$ MeV.
In the application to the neutron star, we select two sets of $(J,\, L,\, K_{\rm sym})$ that give the smallest $\chi^2$ values 
for a given $K_0$ value.
Solving TOV equations, we obtain the mass and radius of the neutron star.
Comparison with neutron star data provides us the ranges of $K_0$, $J$ and $L$ that are most consistent with both nuclear data
and astronomical observation.
In the result, we obtain $K_0$, $J$ and $L$ constrained in ranges 
$K_0 \sim 230-250$ MeV, $J \sim 31-33$ MeV and $L\sim 55-65$ MeV.

In Section II, we introduce the model and explain the strategy of investigation.
Neutron star properties are considered, and reduced ranges of the EoS parameters are obtained in Section III. 
Parameters that are consistent with the neutron star observation are applied to calculating the properties of nuclei in Section IV.
We summarize the work in Section V.

\section{Model and Fitting}

KIDS (Korea:IBS-Daegu-Sungkyunkwan) model provides nuclear energy density functional based on rules for
systematic expansion of the energy of a nucleon in nuclear matter.
In the homogeneous infinite matter, energy per particle is expanded in terms of the Fermi momentum as
\begin{eqnarray}
{\cal E}(\rho,\, \delta) = {\cal T} + \sum_{k=0} (\alpha_k + \beta_k \delta^2) \rho^{1+k/3},
\end{eqnarray}
where ${\cal T}$ is the kinetic energy, and the terms in the summation denote the potential energy originating from the strong interaction.
In the cold nuclear matter $k_F \propto \rho^{1/3}$, so expansion in powers of $\rho^{1/3}$ is equivalent to the expansion 
in terms of the Fermi momentum.
It's been shown that 3 $\alpha_k$'s and 4 $\beta_k$'s are suffcient for a correct description of the EoS of nuclear 
matter at high densities \cite{kidsnuclei2}.

By transforming the KIDS functional to the Skyrme-type potential, one can easily obtain the single-particle potential of the nucleon in nuclei.
In addition to $\alpha_k$ and $\beta_k$ which determine the EoS of infinite nuclear matter,
we have two more parameters ($\zeta$ and $W_0$) in the single particle potential for the nucleon in nuclei.
We have 9 free parameters altogether.
Determination of the 9 parameters is divided into two steps:
At first, we assume specific values of the 7 EoS parameters $\rho_0$, $E_B$, $K_0$, $J$, $L$, $K_{\rm sym}$ and $Q_{\rm sym}$,
and determine 3 $\alpha_k$'s and 4 $\beta_k$'s to reproduce these values.
In the second step, we fit $\zeta$ and $W_0$ to $E/A$ (energy per nucleon) and $R_c$ (charge radius) of $^{40}$Ca, $^{48}$Ca and $^{208}$Pb. 
Agreement to the 6 nuclear data is measured by the $\chi_6^2$ defined by
\begin{equation}
\chi_6^2 \equiv \sum_{n=1}^6 \left(\frac{O^{\rm exp}_n - O^{\rm calc}_n}{O^{\rm exp}_n} \right)^2.
\label{eq:chi6}
\end{equation}

\begin{figure}
\begin{center}
\includegraphics[width=0.6\textwidth]{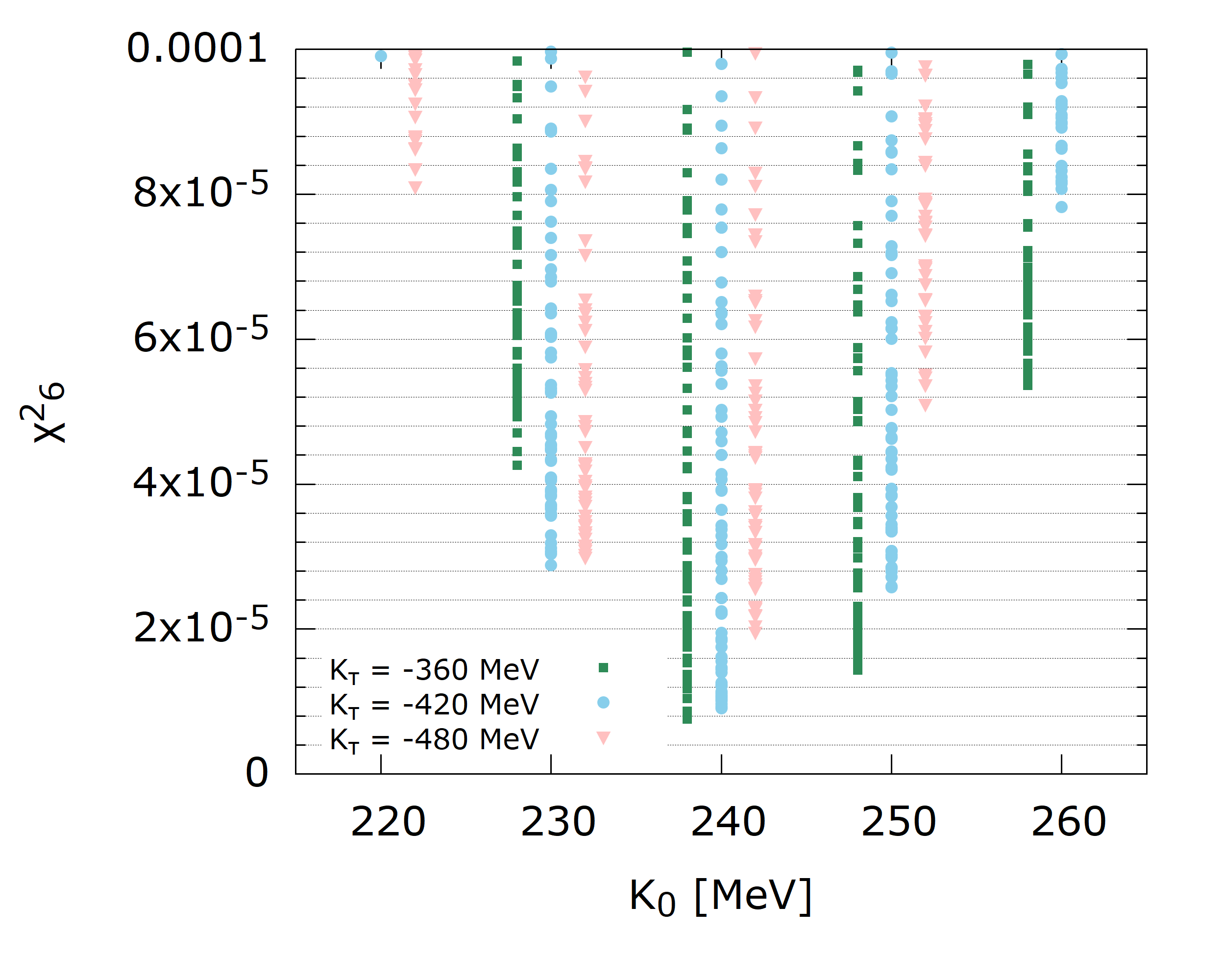}
\caption{$\chi_6^2$ less than $10^{-4}$ for $K_0=$ 220, 230, 240, 250 and 260 MeV.}
\label{fig:chi62}
\end{center}
\end{figure}

Among the 7 EoS parameters, we fix $\rho_0$, $E_B$, $Q_{\rm sym}$ to 0.16 fm$^{-3}$, 16.0 MeV and 650 MeV, respectively,
and investigate the dependence on the 4 EoS parameters $K_0$, $J$, $L$, and $K_{\rm sym}$.
In order to avoid too large parameter space, 4 EoS parameters are restricted to the ranges allowed by either experiments 
or other theories.
For $K_0$, the range is chosen as $220-260$ MeV, and the interval is divided into 10 MeV, i.e. 220, 230, 240, 250 and 260 MeV.
Ranges of $J$ and $L$ are $30-34$ MeV and $40-70$ MeV, respectively, and the interval is assumed equally 1 MeV. 
Instead of $K_{\rm sym}$ we consider $K_\tau$
\begin{equation}
K_\tau \equiv K_{\rm sym} - \left(6 + \frac{Q_0}{K_0} \right) L,
\label{eq:ktau}
\end{equation}
and assume three values $-360$, $-420$ and $-480$ MeV for $K_\tau$.
Total number of the combination of ($K_0$, $J$, $L$, $K_\tau$) is $5 \times 5 \times 30 \times3 =2250$,
and the resulting $\chi_6^2$ ranges from $7.5 \times 10^{-6}$ to $3.2 \times 10^{-4}$.
Among the 2250 results, we plot the results of $\chi_6^2$ less than $10^{-4}$ for each $K_0$ value in Fig.~\ref{fig:chi62}. 
The results are comparable with the SLy4 and the GSkI models which 
give $\chi_6^2$ values $2.24 \times 10^{-4}$ and $6.9 \times 10^{-5}$, respectively.

There are several notable features in Fig.~\ref{fig:chi62}.
First, minimal $\chi_6^2$ is likely to be located in the range $K_0 = 230-250$ MeV.
This range is consistent with, and narrower than the one from experiment $K_0=200-260$ MeV.
Second, with larger $K_0$ values, smaller $\chi_6^2$ is obtained with larger $K_\tau$ values (smaller in magnitude).
Third, dependence on $K_\tau$ is relatively weak in the range $K_0=230 - 250$ MeV.
Interestingly this range agrees with the range of minimum $\chi_6^2$.
Since the nuclei in the fitting are not so asymmetric i.e. $\delta^2 \ll 1$, 
and the densities relevant to the 6 data may be $\rho_0$ and nearby, i.e. $|(\rho-\rho_0)/3\rho_0| \ll 1$,
it would be natural and reasonable to have weak dependence on $K_\tau$.

From the result of fitting to the nuclear data, one can guess that true $K_0$ value may be located in $K_0 = 230-250$ MeV.
However, it is premature to draw a definite conclusion because 
density of nuclei is identically close to $\rho_0$, so the information about the density dependence of the EoS
one can obtain from the nuclear data might be limited. 
In addition, asymmetry $\delta$ of $^{48}$Ca and $^{208}$Pb are 0.17 and 0.21, respectively,
so they are not large enough to magnify the effect of the symmetry energy.
Neutron stars provide environment with more wider density range and the large neutron-proton asymmetry. 
Not abandoning the possibility of small or large $K_0$ value, we include $K_0=220$ and 260 MeV in the consideration of the neutron star.
For each $K_0$ value, ($J$, $L$, $K_\tau$) values are chosen from the two smallest $\chi_6^2$ values.
Equation of state parameters thus determined are summarized in Tab.~\ref{tab:kjl}.
Ranges of $J$ and $L$ in Tab.~\ref{tab:kjl} are $30-34$ MeV and $40-70$ MeV, respectively.
These ranges are similar to the experimental ranges adopted in \cite{dutra}.

\begin{table}
\begin{center}
\begin{tabular} {|ccc|}\hline
$K_0$ & $(J,\, L\, K_\tau)$ & $\chi^2 (\times 10^{-5})$ \\ \hline\hline
220 & (33, 50, $-480$) & 9.45 \\
     & (34, 63, $-480$) & 8.61 \\ \hline
230 & (33, 66, $-420$) & 3.04 \\
  & (33, 52, $-480$) & 3.01 \\ \hline
240 & (32, 68, $-360$) & 0.75 \\
  & (32, 58, $-420$) & 0.89 \\ \hline
250 & (30, 41, $-360$) & 1.50 \\
 & (31, 58, $-360$) & 1.43 \\ \hline
260 & (30, 47, $-360$) & 5.55 \\
 & (31, 63, $-360$) & 6.03 \\ \hline
\end{tabular}
\end{center}
\caption{EoS parameters ($J$, $L$, $K_\tau$) giving the two smallest $\chi_6^2$ values for each $K_0$ value.}
\label{tab:kjl}
\end{table}

\section{Neutron star}

In 1990's when data of mass and radius were not sufficient, mass observation played a major role in
constraining the nuclear EoS at high densities.
For various reasons, broadly agreed maximum mass was about $1.5 M_\odot$.
However masses close to or larger than $2M_\odot$ have begun to be reported in the 21st century,
and now the general consensus is that the maximum mass of the neutron star may be $2 M_\odot$ or more.
In the core of $2M_\odot$ neutron star, density is large enough that the nucleons are overlapping to each other.
It is very likely that non-nucleonic state will emerge and exist in the core of $2M_\odot$ neutron stars,
so the EoS at the core is hard to determine definitely because of large uncertainties from various sources.

On the other hand, density at the center of a $1.4M_\odot$ neutron star does not exceed $3 \rho_0$ regardless of the siffness of EoS.
At densities smaller than $3\rho_0$, uncertainty due to the non-nucleonic degrees of freedom can still exist, but their effects
are not significant or critical.
For example, mass and radius of neutron stars with $\Lambda$ hyperons were calculated with EoSs over a wide range 
of stiffness for the nucleon-nucleon ($NN$), $N\Lambda$ and $\Lambda\Lambda$ interactions \cite{hyperon2015}.
Maximum mass and the mass-radius curves depend on the $N\Lambda$ and $\Lambda\Lambda$ interactions critically, 
but the mass-radius relation of the canonical neutron star is hardly affected by the existence of $\Lambda$ hyperons and the stiffness of 
their interactions with $N$ and $\Lambda$.
Therefore $1.4M_\odot$ neutron star provides a good laboratory in which we can study the EoS of nucleonic matter at densities
higher than $\rho_0$.

\begin{figure}
\begin{center}
\includegraphics[width=0.65\textwidth]{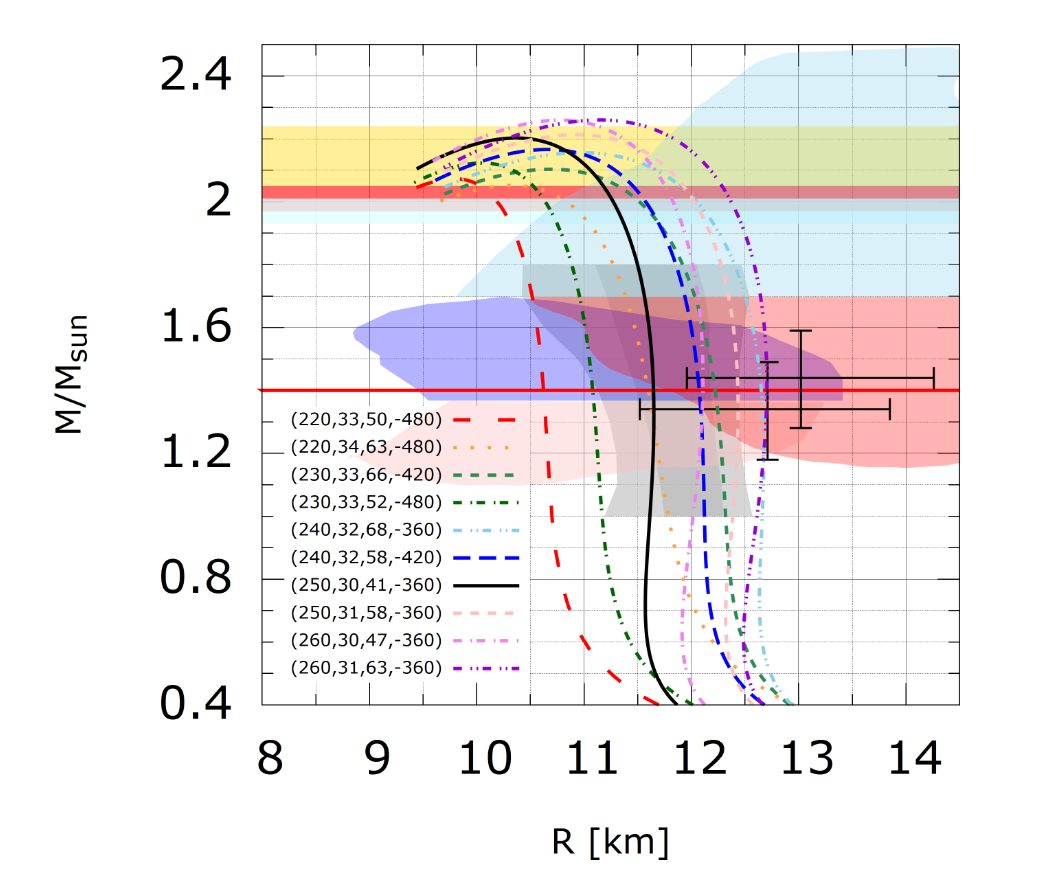}
\end{center}
\caption{Neutron star mass and radius relations with $K_0$, $J$, $L$ and $K_\tau$ selected in Tab. \ref{tab:kjl}.}
\label{fig:ns}
\end{figure}

Now more ample and exact data of mass and radius of neutron stars whose mass is close to $1.4M_\odot$ are available.
We collect the state-of-the-art data of the neutron neutron star in Fig. \ref{fig:ns}.
Horizontal bands around $2 M_\odot$ exhibit the data of large masses \cite{largemass1, largemass2, largemass3}.
Gray zone represents the range obtained from the low mass X-ray binary (LMXB) data \cite{lmxb},
error bars denote the analysis of a soft X-ray source observed in the NICER project \cite{nicer1, nicer2},
and the red and the blue regions in the shape of wings correspond to the result of GW170817 (lower small wings) \cite{gw170817}
and GW190425 (upper large wings) \cite{gw190425}.
In the case of NICER two analyses were performed independently for a single object, so the two results are shown in the figure.
A red horizontal line is drawn at mass $1.4M_\odot$ for reference.
Observation data spread broadly on the mass-radius plane, but there is a narrow band in which all the data overalp.
For the mass $1.4M_\odot$, the commonly overlapping region restricts the radius in the range $11.8-12.5$ km.

Ten candidates for the EoS parameters $(K_0,\, J,\, L,\, K_\tau)$ also show widely spread behavior.
At first compared to the LMXB data at the canonical mass (red line), 4 EoSs are incompatible:
2 soft EoSs corresponding to (220, 33, 50, $-480$), (230, 33, 52, $-480$), and 2 stiff EoSs with (240, 32, 68, $-360$), (260, 31, 63, $-360$).
It is notable that $K_0$ and $K_\tau$ are small simultaneously for the 2 soft EoSs, and they are large simultaneously for the 2 stiff EoSs.
On the other hand, we have 4 EoSs that fall into the common range of LMXB, NICER and GW for $M=1.4M_\odot$:
$(K_0,\, J,\, L,\, K_\tau) = (230,\, 33,\, 66,\, -420),\, (240,\, 32,\, 58,\, -420),\, (250,\, 31,\, 58,\, -360),$ and 
$(260,\, 30,\, 47,\, -360)$. 
There are several interesting feature.
First, the smallest $K_0$ and $K_\tau$ values, 220 MeV and $-480$ MeV are ruled out.
This provides the lower bound of these parameters.
Second, similar to the 2 soft and 2 stiff EoSs incompatible with the LMXB data, 
smaller $K_0$ values are grouped with smaller $K_\tau$ values, and vice versa.
On the other hand, $J$ and $L$ values are smaller when $K_0$ and $K_\tau$ are large.
This means $J$ and $L$ contribute to EoS oppositely to $K_0$ and $K_\tau$, i.e.
if $J$ and $L$ soften the EoS, the latter two parameters make it stiff.
As a result, EoS is adjusted to remain in ranges consistent with data.
Most essential aspect of the result is the range of each EoS parameters.
If the consistency with neutron star observation is accounted, acceptable ranges could be specified as
\begin{eqnarray}
K_0 \sim 230-260,\, J \sim 30-33,\, L \sim 45-65
\label{eq:wide}
\end{eqnarray}
all in the unit of MeV. 
On the other hand, if we take into account the consistency with the nuclear data in addition to the nuclear star observation,
$K_0 = 260$ MeV could be excluded because $\chi^2_6$ is much larger than the other $K_0$ values. Then we obtain narrower ranges
\begin{eqnarray}
K_0 \sim 230-250,\, J \sim 31-33,\, L \sim 55-65.
\label{eq:narrow}
\end{eqnarray}
The ranges are reduced significantly compared to those constrained from experiment \cite{dutra},
and thus they are expected to give EoSs less uncertain from sub to supra saturation densities.
Compared to a recent work \cite{arxiv2020} in which $K_0$ is fixed to 240 MeV and symmetry energy parameters are
fit to 13 nuclear data, $J$ is slightly larger in this work, but $L$ and $K_\tau$ ranges are almost identical.
With the newly constrained ranges of the EoS parameters,
we can obtain uncertainty range of the extreme properties of nuclei such as the neutron skip thickness or
the position of the neutron drip line. Consideration about the finite nuclei follows in the next section.

\section{Nuclear property}

Ranges of the EoS parameters in Eq. (\ref{eq:wide}) are determined using 6 nuclear data of $^{40, 48}$Ca, $^{208}$Pb,
and the neutron star mass and radius.
Since we are aiming at a unified description of both nuclear matter and nuclei,
we should check the predictive power of our approach for the nuclear properties.
Top prior quantity among numerous nuclear properties might be the binding energy.

\begin{table}
\begin{center}
\begin{tabular}{|c|cccccc|}\hline
 & $^{16}$O & $^{40}$Ca & $^{48}$Ca & $^{90}$Zr & $^{132}$Sn & $^{208}$Pb \\ \hline\hline
Exp. & 7.976            & 8.551  & 8.667 & 8.710 & 8.355 & 7.868   \\ \hline
A & 7.946 (0.38) & 8.564 ($-0.16$) & 8.680 ($-0.15$) & 8.681 (0.33) &  8.378 ($-0.28$) & 7.871 ($-0.04$)  \\ \hline
B & 7.940 (0.45) & 8.555 ($-0.04$) & 8.673 ($-0.07$) & 8.676 (0.39) & 8.377 ($-0.26$) & 7.869 ($-0.01$)  \\ \hline
C & 7.935 (0.51) & 8.546 (0.06) & 8.667 ($-0.01$) & 8.678 (0.37) & 8.375 ($-0.24$) & 7.868 (0) \\ \hline
D & 7.933 (0.54) & 8.538 (0.15) & 8.653 (0.17) & 8.674 (0.42) & 8.374 ($-0.22$) & 7.867 (0.02)  \\ \hline  
\end{tabular}
\end{center}
\caption{Binding energy per nucleon in MeV for $(K_0,\, J,\, L,\, K_\tau)$ values satsifying the neutron star mass-radius observation.
Set A, B, C and D correspond to $(230, 33, 66, -420)$, $(240, 32, 58, -420)$, $(250, 31, 58, -360)$, and $(260, 30, 47,-360)$, respectively.
Numbers in parenthesis denote the difference from experiment in \%.}
\label{tab:be}
\end{table}

Table \ref{tab:be} displays the binding energy per nucleon for standard spherical magic nuclei.
Numbers in the parentheses denote the difference from experiment in units of \%.
Energies of $^{40, 48}$Ca and $^{208}$Pb are used in fitting the 2 Skyrme force parameters that are coefficients of the 
grandient and spin-orbit terms, so it is natural to have good reproduction of these data. 
Energies of $^{16}$O, $^{90}$Zr and $^{132}$Sn are, on the other hand, predictions of the model.
The predictions agree well with experiment, giving difference from experiment at the order of 0.5\% or less.
Another positive aspect of the result is that the predicted energy values are independent of the EoS parameters
($K_0$, $J$, $L$, $K_{\rm sym}$).
This indpendence implies that the 4 EoS parameter sets have equal predictive power for the binding energy of magic spherical nuclei.


\begin{figure}
\begin{center}
\includegraphics[width=0.8\textwidth]{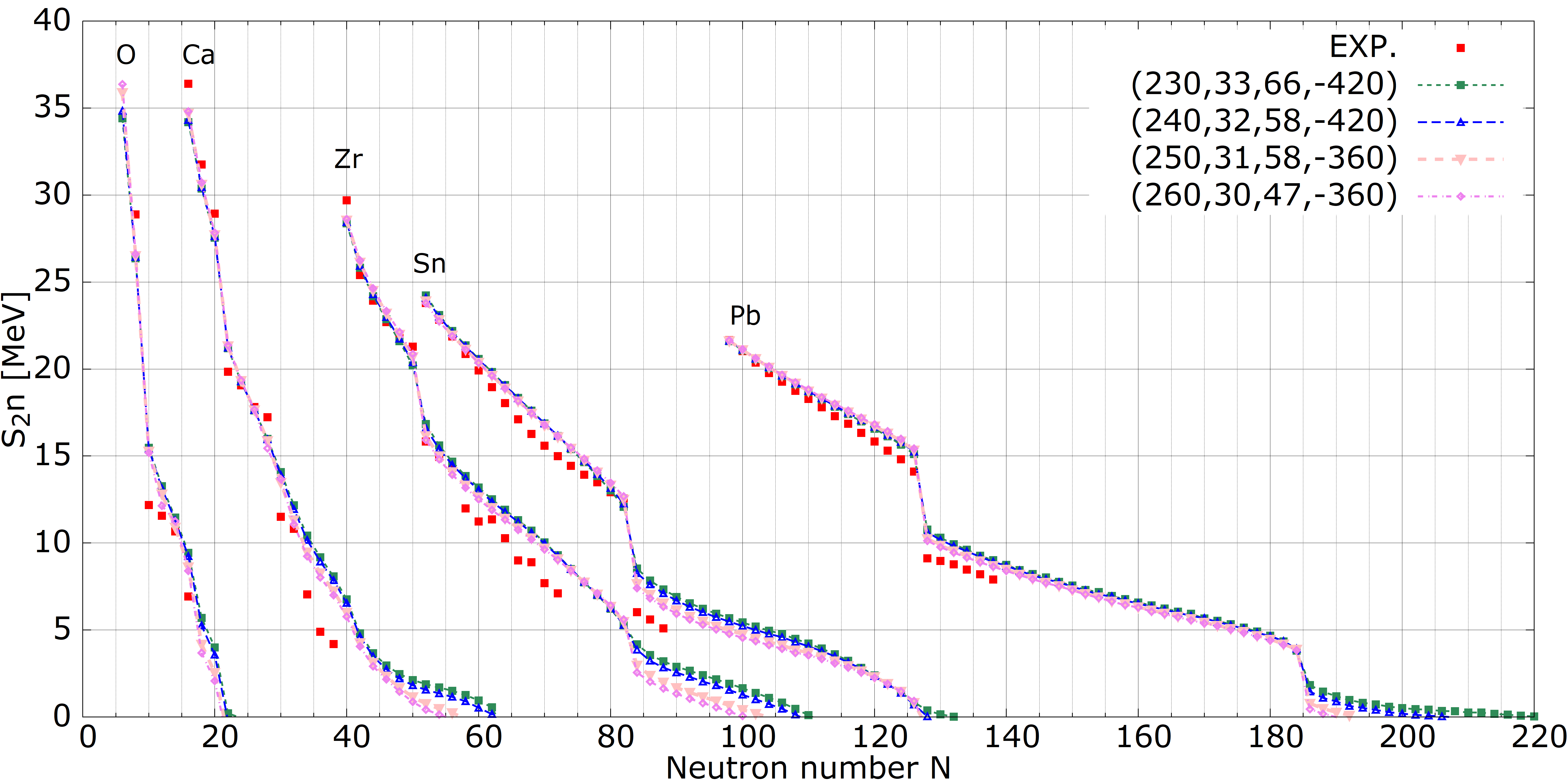}
\end{center}
\caption{Two neutron separation energies for the O, Ca, Sn and Pb isotopic chains 
for the four sets of $(K_0, J, L, K_\tau)$ in Tab. \ref{tab:be}.}
\label{fig:s2n}
\end{figure}

One of the main goal of the work is to have reliable predictions about the objects in extreme conditions in which density is
much larger or smaller than the saturation density, or proton-neutron asymmetry approaches to 1.
Neutron stars are good laboratory to explore these effects, but nuclei close to the neutron drip line also provide a nice test ground.
In Fig. \ref{fig:s2n}, we show the result of two neutron separation energy $S_{2n}$ of O, Ca, Zr, Sn and Pb isotopes 
calculated with 4 EoS parameter sets in Tab. \ref{tab:be}.
Experimental data are included with red squares.
In order to take into account the effect of continuum state correctly, we use the Hartree-Fock-Bogolyubov theory.
All the isotopes are assumed spherical, and pairing force is incorporated in the contact interaction form
\begin{eqnarray}
V_{\rm pair} = t'_0 \left( 1- \frac{\rho}{2 \rho_0} \right) \delta(\vec{r}_1 - \vec{r}_2).
\label{eq:pairing}
\end{eqnarray} 
Parameters in the pairing force are fit to the mean neutron gap of $^{120}$Sn, 1.392 MeV.

Agreement to the experiment is not as good as the energies of the magic nuclei in Tab. \ref{tab:be}.
More investigation is necessary for a better reproduction of the data.
For example, the disagreement can originate from the way to treat the pairing force. 
Non-spherical shape of the isotopes could contribute to the discrepancy.
Nevertheless, there are a few points that are worthy to be discussed.
For the oxigen, results of the 4 EoS parameter sets agree to each other fairly well, but there are huge differences from experiment at $N=8$ and $N=10$.
Position of the neutron drip line is predicted identically at $N=22$.
For the calcium, agreement to experiement is reasonable for $N$ less than 30, but it becomes dramatically bad for $N>30$.
For $N$ between 30 and 40, however, the 4 EoS sets predict similar results,
and for $N>40$, predictions are divided into two branches.
Sets with $K_\tau=-420$ MeV give $S_{2n}$ larger than those of $K_\tau=-360$ MeV sets.
Consequently neutron drip lines differ significantly, $N=54$ for $K_\tau=-360$ MeV and $N=62$ for $K_\tau=-420$ MeV.
For the zirconium, agreement to experiment becomes worse for $N>40$.
Four EoS parameter sets agree to each other well for $N$ less than 82, but the prediction splits into two groups for $N>82$.
Similar behavior is observed for the tin isotopes.
For the lead, the same thing happens at another magic number $N=184$.
Neutron drip of the Pb isotopes spread over a huge range of the neutron number, from $N=186$ to $N=220$.

\section{summary}

This work was motivated by the observation that the EoS of nuclear matter which is crucial in understanding the properties of
extremely neutron-rich nuclei and the neutron star is not determined precisely enough yet.
By reducing the uncertainty in the EoS parameters $K_0$, $J$, $L$, and $K_{\rm sym}$, it may be feasible to have more
exact knowledge about the state of matter at extreme conditions.
KIDS functional theory provides a framework adequate for this analysis.
Four EoS parameters are determined to best reproduce the binding energy and the charge radius of $^{40, 48}$Ca and $^{208}$Pb,
and the radius of $1.4M_\odot$ neutron stars determined from various sources and probes.
As a result we obtain the ranges $K_0 \sim 230-250$ MeV, $J \sim 31-33$ MeV, and $L\sim 55-65$ MeV.
As for $K_\tau$, we used three values $-360$, $-420$ and $-480$ MeV, and find that $-480$ MeV makes the EoS too soft 
to satisfy the neutron star observation.
Ranges of $J$ and $L$ are consistent with the ranges in a recent work \cite{arxiv2020} in which $K_0$ is fixed to 240 MeV
and the nuclear data used in the fitting are chosen differently from this work.
EoS parameters thus determined are used to calculate the properties of nuclei.
For the binding energies of spherical magic nuclei $^{16}$O, $^{90}$Zr and $^{132}$Sn, all the 4 EoS parameter sets 
consistent with the constraint from the neutron star give results agreeing to nuclear data within the errors less than 0.5\%.
Two-neutron separation energy, on the other hand, shows substantial discrepancy with experiment, so it needs more study 
to understand the origin of the discrepancy and make the theoretical prediction better.

From the analysis of this work, it has been shown that the uncertainties of $J$ and $L$ are reduced to the ranges 
$\pm 1$ MeV and $\pm 5$ MeV, respectively.
The result of two-neutron separation energy shows that the neutron drip line can be sensitive to the value of $K_\tau$.
We can write an approximate form of Eq. (\ref{eq:ktau}) as
\begin{equation}
K_\tau \simeq K_{\rm sym} - 5L.
\end{equation}
If $L$ is determined with errors $\pm 5$ MeV, maximum difference 10 MeV can give about 50 MeV uncertainty in $K_\tau$.
For a better decision of $K_\tau$, it might be necessary to constrain $L$ in regions narrower than $\pm 5$ MeV.
Aside from the uncetainty of the EoS parameters, it is important to have more correct description of the exotic nuclei.
There are many things to consider, e.g. refined treatment of the pairing force, the effect of deformation, and etc.
Some of these subjects are under way. 

\section*{Acknowledgments}
This work was supported by the Daegu University Research Grant 2018.

\end{document}